# Linear and Nonlinear Fano Resonance on two-dimensional Magnetic Metamaterials

H. Liu,[1,2] G. X. Li,[1,*] K. F. Li,[1] S. M. Chen,[1] S. N. Zhu,[2]

C. T. Chan,[3] and K. W. Cheah[1]

[1]*Department of Physics, Hong Kong Baptist University, Hong Kong, China*
[2]*Department of Physics, National Laboratory of Solid State Microstructures, Nanjing University, Nanjing 210093, China*
[3]*Department of Physics and Nano Science and Technology Program , Hong Kong University of Science and Technology, Clearwater Bay, Hong Kong, China*

We demonstrate that both linear and nonlinear Fano resonances can be realized on two dimensional magnetic metamaterials. The Fano resonance comes from the interference between localized magnetic plasmon resonance and propagating surface plasmon polaritons. When studying the linear optical response of the metamaterial structure, this interference phenomenon was observed in the ellipsometric spectrum. By finely tailoring the geometrical parameters of the magnetic metamaterial device, the nonlinear Fano response was tuned to a near-infrared wavelength (1.61–1.8 μm) of femtosecond pump laser, and Fano-type modulation of the third harmonic generation was found and agrees well with our theoretical model.

# I. Introduction

Recent studies on nonlinear optical plasmonic nanostructures show that the efficiency of nonlinear optical process can be efficiently enhanced by the strongly confined plasmonic excitations.[1-16] By using enhanced optical nonlinearity in metallic nanostructures, plasmonic/metamaterial devices have also been used for ultrafast optical switching application.[17-19] In field of nonlinear optics, while the second order process was extensively studied on various plasmonic/metamaterial nanostructures,[2-10] the third order nonlinear optical phenomena of metallic nanostructure are attracting much more attentions. For example, surface enhanced four-wave mixing on gold plasmonic nanostructure,[11,12,14] plasmon enhanced third harmonic generation[15,16] were demonstrated by virtue of the strong field localization due to plasmonic excitations. The strong field localization discussed usually refers to the electric field. However, the magnetic field also significantly contributes to the enhancement of electric field under resonant condition. In one pioneering work of J. Pendry,[20] magnetic resonance induced field localization was proposed to enhance the efficiency of near linear optical process. For magnetic plasmon resonance mode, the energy of incident light is absorbed into the equivalent LC circuit and strongly localized in a ultra small volume of the equivalent capacitor. The strong field localization makes MPR nanostructure a good candidate for surface enhanced nonlinear optical process. Using this optical property, enhanced nonlinear optical process in magnetic metamaterials i.e.THG,[21] nanolaser,[22,23] and spaser[24,25] have been reported in past several years.

In complex plasmonic systems, usually, two or more resonant modes exist. The interference between these plasmonic modes can result in strong Fano resonance.[25] Linear optical property of Fano resonance in plasmonic nanostructures has been extensively studied recently[26-33]. Integration of Fano resonance into nonlinear optics is expected to offer more optical functionalities, for example, enhanced nonlinear optical response of the free carriers in metamaterial/carbon nanotube composite[30] and fishnet metamaterial[34, 35] has been utilized to fabricate ultrafast optical switch.

In this work, we demonstrate that third harmonic generation, one kind of radiative modes in third order nonlinear optical process, can be spectrally modulated using nonlinear Fano-resonant magnetic metamaterials. The device is a tri-layer nanostructure, consisting of gold slab/ITO/two dimensional gold nanodisc. This Fano-resonant metamaterial supports both localized MPR mode and propagating SPP mode. By finely tuning the interference between the localized MPR and propagating SPP modes at near infrared wavelength, the surface polarization of pump laser on metamaterial has Fano-resonant modulation. Spectral resolved third harmonic generation on this metal/dielectric hybrid nanostructure was demonstrated to show a Fano-resonant modulation. This indicates that third order and other nonlinear optical processes can be tailored by engineering the coupling of the MPR and SPP mode, the interference between them provides us a more flexible way to design nonlinear optical nano-devices.

## II.     Device Fabrication

The cross-section of unit cell of nanostructure is shown in Fig. 1(a), and the unit cell is composed of a gold nanodoisc fabricated on top of ITO/gold slab. Two samples (A and B) were fabricated using e-beam lithography, thin film deposition, and metal lift-off process. The SEM pictures of the two samples are shown in Fig. 1(b) and Fig. 1(c). The thicknesses of gold nanodisc, ITO and gold slab are 50 nm thick respectively. The disc diameter of A is 320 nm with period of 600 nm, while for sample B, the diameter of disc is 360 nm with the period of 910 nm.

## III.     Optical Characterization

The linear optical property of this metamaterial was characterized using Fourier Transform Infrared spectrometer (FTIR) and spectroscopic ellipsometer. In FTIR experiment, the reflection spectra of unpolarized light (shown in Fig. 2(a) on sample A and B were measured using normal incidence setup. A thick gold film (100 nm thick) coated silicon substrate was used as reference sample. The relative reflection spectra are calculated from the ratio of reflected intensity on the nanostructure and gold reference. The magnetic resonant dips were found at 1.51 μm for sample A (line with triangle symbol) and 1.62 μm for sample B (line with circle symbol). At the resonant wavelength, the induced currents in disc and slab are anti-parallel, which correspond to MPR modes as reported before.[23, 36, 37] In ellipsometry measurement, the reflection coefficient ratio of TM and TE waves $\eta = E_{TM}/E_{TE}$ was retrieved using rotating polarizer technique. In Fig. 2(b), spectral resolved $\eta$ is measured at the incident angle: $\theta=52^o$. For sample A, pure MRR mode still exists and it has a little red shift compared to the FTIR result (Fig. 2(a)). However, for sample B, a Fano-type optical response at MPR wavelength is observed. This interesting phenomenon means that the resonance in sample B is not a pure MPR mode. The angle resolved ellipsometry measurement (incident angle $\theta$ was tuned from $48^o$ to $56^o$) was then used to study the mechanism behind this phenomenon. For sample A, magnetic resonance mode is not sensitive to the incident angles and the reflection dip almost remains at the same wavelength as shown in Fig. 2(a). Angle resolved reflection ellipticity of sample B is shown in Figs. 3(a-c) (dashed line). The Fano resonance is very sensitive to the incident angle and shifts to longer wavelength when increasing incident angle.

## IV. Theoretical Analysis of Fano Resonance

As MPR is a kind of localized mode with a definite resonance frequency and it should be not sensitive to the incident angle.[36] From our previous study on plasmonic ellipsometry,[38] the angle dependent shift of Fano peak on sample B is related to the excitation of propagating surface plasmon polaritons. The excitation of propagating SPP wave is governed by the following momentum conservation condition: $k_{//}+k'_{spp}-G'=0$ (Fig. 4).[39] Here, $k_{//}=(\omega/c)sin\theta$ is the x component of vacuum wavevector; $k_{spp}=k'_{spp}+ik''_{spp}$ is the wave vector of SPP waves, with the imaginary part $k''_{spp}$ representing the ohmic loss of SPP wave; $G=2\pi/period+iG''$ is the first order of reciprocal vector, and the $G''$ is the radiative loss. For sample B (with period of 910 nm), the wavelength of first order SPP is ~1.63 μm at incident angle:

$\theta=52°$, which is very close to the MPR mode with wavelength of ~ 1.62 μm. Thus both SPP and MPR modes can be excited (see Fig. 4), and the interference between MPR and SPP mode shows a Fano type shape in reflection spectra (Fig. 3). But for sample A, as its period (600 nm) is too small, the first order SPP (~ 1.08 μm at $\theta=52°$) is far away from the MPR wavelength (1.51 μm), that is why there is no similar Fano-type resonance.

When propagating surface plasmon polaritons are excited, the reflected TM waves shows a resonant dip at the resonant frequency and the reflected TE wave is slowly varying on the spectrum. The plasmonic excitation can be characterized from the spectrum of ellipticity ($\eta$).[33] Around resonant wavelength of surface plasmon polaritons, the TM polarized reflection spectra for SPP is defined as:[39]

$$r_{spp}=r_{p0}\left(1-\frac{2iG''}{k_{//}+k_{spp}-G}\right). \quad (1)$$

Here, $r_{p0}=(n_{gold}cos\theta-cos\theta_{gold})/(n_{gold}cos\theta+cos\theta_{gold})$ is the reflection coefficient on air/gold surface ($\theta_{gold}$ is the refractive angle in gold film); $G_0=(2\pi/p)(1+i\Gamma)=G'+iG''$ is the reciprocal vector and $\Gamma$ is radiative damping rate of surface plasmon polariton (estimated from Fig. 2(b)); $k_{spp}(\omega)=k_0(\omega)\sqrt{\varepsilon_{gold}\varepsilon_{air}/(\varepsilon_{gold}+\varepsilon_{air})}$ is wave vector of SPP wave at interface between gold and air.

As was reported before,[37] the nanostructure in Fig. 1(a) is regarded as an equivalent LC circuit and described by Lagrangian as $L=\frac{L}{2}\dot{Q}^2-\frac{1}{2C}Q^2$. Here, L and C are the inductance and capacitance of the LC circuit. Q is the net charge in the capacitor and $\dot{Q}$ is the induced current in the inductance. In the presence of ohmic dissipation and an external driving field, the Euler-Lagrange equation can be written as: $\frac{d}{dt}\left(\frac{\partial L}{\partial \dot{Q}}\right)-\frac{\partial L}{\partial Q}=-\frac{\partial R}{\partial \dot{Q}}+e.m.f.$, where $R=R_{eff}\dot{Q}^2/2$ ($R_{eff}$ is the effective resistance in the system) and the e.m.f. (for normal incident $B_0$ field) is given by: $e.m.f.=-\frac{dB_0}{dt}S_{eff}$ ($S_{eff}$ is the effective cross-section area between patch and slab). If we define an effective magnetic dipole as: $m=S_{eff}\cdot\dot{Q}=\alpha_m\cdot B_0$ ($\alpha_m$ is the effective magnetic polarizability of magnetic resonator), the Euler-Lagrange equation has the following solution: $\alpha_m=\frac{S_{eff}^2}{L}\frac{\omega^2}{(\omega^2-\omega_0^2)+i\gamma\omega}$. At MPR frequency ($\omega_0$ is estimated from Fig. 2(b)), the electromagnetic energy is absorbed by the magnetic resonators and finally converted to heat loss. For a pure MPR mode, a Lorenz-shape absorption dip will be observed in the reflected spectrum. Based on the above Lagrangian model, the reflection coefficient of MRC can be described by $r_{MPR}=(1-C_{MPR})$, where the absorption coefficient is $C_{MPR}=f_{MPR}\frac{i\gamma_m\omega}{(\omega^2-\omega_0^2)+i\gamma_m\omega}$,[40] and $f_{MPR}$ is the light coupling efficiency of magnetic resonators. For TM polarization, as the magnetic field is always perpendicular to the incident plane. The electromotive force is not sensitive to the incident angle, so the reflection coefficient can be given by $r_{MPR\_TM}=r_{p0}(1-C_{MPR})$. In comparison, $r_{MPR\_TE}=r_{s0}(1-C_{MPR}cos\theta)$ is used to calculate the reflection coefficient for TE polarization, as the electromotive force is dependent on the incident angle $\theta$ as: $e.m.f.=-\frac{dB_0}{dt}S_{eff}cos(\theta)$ in this case.

For TM wave, the total reflection coefficient for TM wave is given by the interference of SPP and MPR modes:

$$r_{TM}=(f_{spp}r_{spp}+(1-f_{spp})r_{mpr\_TM}), \qquad (2)$$

where $f_{spp}$ and $(1-f_{spp})$ are the weight factors to describe the ratio of SPP and TM polarized MPP

mode respectively. For TE wave, the modulated reflection spectrum is mainly due to the absorption of TE polarized MPR mode, the approximated reflection coefficient is:

$$r_{TE} = r_{mpr\_TE}, \tag{3}$$

To compare the theoretical calculation with experimental results, a modification factor $f_{ellip}=1.583$ is introduced. As shown in Figs. 3(d-f) (solid line), the theoretical ratio of reflection coefficients ($\eta = f_{ellip} r_{TM} / r_{TE}$) for TM and TE polarization are shown ($f_{spp}=0.16$ and $f_{mpr}=1.1$ are chosen). It can be found that Fano resonances appear in the calculated results and agree with experimental measurement (Figs. 3(a-c)).

## V. Nonlinear Optical Experiment

Nonlinear optical property of Fano resonant metamaterial was studied by using third harmonic generation process. A femtosecond laser pumped optical parametric amplifier (OPA, TOPAS-C) system was used in this experiment. The output wavelength of OPA can be tuned from 1.61 μm to 1.8 μm with repetition frequency of 1 kHz and pulse duration ~100 femtosecond. As is shown in Fig. 5, the horizontally polarized (electric field parallel to incident plane) pump laser was incident at angle of 52°. After being focused by a lens with focal length of 150 mm, the near infrared laser pulse was used to pump the metamaterial nanostructure. The laser spot size on the nanostructure is about 120 μm by 100 μm, which is less than the total area size of the nanostructure.

The radiation direction of third harmonic generation is determined by the momentum conservation condition, which is given by the following equation [9]:

$$k_0(3\omega)\sin(f)=3k_0(\omega)\sin(\theta)+m\frac{2\pi}{p_x}+n\frac{2\pi}{p_y}$$

where $f$ is the radiation angle of third harmonic generation to the surface normal of the nanostructures, $m$ and $n$ represent the diffraction orders. For the in-plane first order ($m=-1, n=0$) emission of THG, radiation angle ($f$) ranges from 11.4 to 7.39 degree while excitation wavelength is tuned from 1610 nm to 1800 nm. THG signal was collected by silicon based CCD detector and spectrometer (Ocean Optics 4000) after filtering the fundamental wave. The output power was measured by a photodiode (Newport 818-UV). Two linear polarizers were used to control the power and polarization of pump laser, the Newport power detector (818-IR) was used to monitor the power variation of pump laser (NIR) from the side beam of beam splitter.

An imaging system was set up to monitor the position of nanostructure and pump laser. White light source (Xenon lamp) was used to illuminate the nanostructure from the quartz substrate. In bright field mode (Fig. 6a and Fig. 6b), it can be found that both the radiation spot (marked by dashed line) of third harmonic generation and nanostructure can be imaged by CCD camera. After turning off the illumination light, the dark field image of third harmonic generation (Fig. 6c and Fig. 6d) was realized. The spectrum of third harmonic generation was measured using the Ocean Optics spectrometer after filtering the pump laser. In Fig. 7, the THG spectra on sample A are given with fundamental wavelength of 1620 nm and 1640 nm. Bandwidth of the THG peak is about 9 nm. The relative efficiency of THG (intensity ratio of THG and pump laser) on sample A and B are given in Fig. 8. For sample A, which has resonant wavelength at ~ 1.51 μm, the third harmonic generation efficiency decreases when pump wavelength is tuned from 1.61 to 1.8 μm (Fig. 8(a)). However, the spectrum of third harmonic generation on sample B is quite different, in which a Fano-resonant modulation was obtained (Fig. 8(b)).

To elucidate the third harmonic generation on the metamaterial devices, a nonlinear modulation factor of THG is defined: $\eta_{THG}=\eta_0+\eta_{MPC}$, where $\eta_0$ and $\eta_{MPC}$ come from intrinsic and field enhancement contribution. The surface nonlinear polarization on the metallic nanostructure can be written as: $\vec{P}^{(3)}_{SNL}=\chi^{(3)}\mathbf{M}(\eta_{THG}\vec{E}_0(\omega))^3$, and $\vec{E}_0$ is the electric field of pump laser.[36] The intrinsic THG contribution ($\eta_0$) in this experiment is attributed to $\chi^{(3)}$ of gold and ITO. As reported in previous work,[15] the gold plasmonic nanostructure shows highly efficient THG radiation at the near infrared wavelength (1.61 to 1.8 μm). Most interestingly is that the three pumping photons in THG process do not induce the interband transition (from 5d to 6sp) in gold, as the band gap of 5d-6sp for gold is about 2.4 eV (~ 520 nm). The excited electrons in gold are pumped to a virtual state near to Fermi level, the dephasing time of electrons at this state should be an ultrafast process, and this explains the efficient THG radiation. In addition, the third order susceptibility of ITO should also be taken into account,[42] especially for magnetic resonant nanostructure in which the energy of pump laser is also strongly localized in ITO spacer layer.

For sample A, field localization mainly comes from the magnetic plasmon resonance, the modulation factor of electric field can be described by $\eta_{MPC}=Q_{MPR\_A}|C_{MPR\_A}|$. For sample B, $\eta_{MPC}$ comes from the contribution of interference between MPR and SPP modes, it can be written as: $\eta_{MPC}=|Q_{MPR\_B}C_{MPR\_B}+Q_{SPP}C_{SPP}|$,

where $Q_{MPR}$ and $Q_{SPP}$ are the weight factors of MPR and SPP modes (ratio=7 is chosen in calculation). Assuming that ITO and gold have similar third order susceptibility,[42] radiation intensity of THG has cubic relation with power of pump laser, thus we have $I_{THG} \sim |\vec{P}_{SNL}|^2 \sim |\eta_{MPC}^2 \vec{E}_0^2|^3$. The calculated results (line with triangle symbol) agree with the measured ones (line with circle symbol). Based on above theoretical analysis, we know that the Fano type THG comes from modulated surface nonlinear polarization of pump laser, indicating that the nonlinear polarizations from the surface plasmon polaritons and MPR modes have an interference effect. This kind of interference can bring more modulation functions. By tuning the resonant frequency and the phase delay between SPP and MPR modes, the plasmonic excitation around Fano resonant wavelength can be switched from the localized magnetic resonance mode to propagating surface plasmon mode, or vice versa.

## VI. Conclusions

In conclusion, we designed a two-dimensional magnetic metamaterial. Both localized MPR and propagating SPP modes can be excited in such a system. Through the interference between MPR and SPP excitations, linear Fano effect is found in the ellipsometric spectra. The nonlinear optical process on the metamaterial device was studied through THG experiment. By modulating the nonlinear polarization through the interference of MPR and SPP modes, Fano-type THG was obtained in our structure. The integration of Fano resonance into nonlinear optical MPR device will provides many more opportunities for realizing nonlinear optical nano-device.

Although we only studied THG here, such kind of nonlinear Fano resonant effect can also be applied to other nonlinear optical processes, such as high harmonic generation, surface enhanced Raman, four-wave mixing and etc.

## ACKNOWLEDGMENTS

G. X. would like to thank Dr. Nelson Li and Prof. J. N. Wang for their support on device fabrication. Device fabrication was supported by University Grant Council with grant SEG_HKUST10. H. L. was financially supported by K. C. Wong Education Foundation, National Natural Science Foundation of China (No. 11021403, 11074119, 10874081, 60990320 and 11004102), and by the National Key Projects for Basic Researches of China (No. 2010CB630703, 2012CB921501 and 2012CB933501). G. X. and K. W. would like to thank Prof. John Pendry, Dr. T. Li and Dr. S. M. Wang for fruitful discussions.

*sunview1981@hotmail.com.

**FIGURE CAPTIONS**

**FIG. 1.** (Color online) (a) Cross-section of the nanostructure: gold circular disk/ITO layer/gold slab; (b) and (c) SEM picture of MPC samples: gold disk array/ITO layer/gold slab/Cr layer/quartz. Thickness of gold/ITO/gold/Cr is 50 nm/50 nm/50 nm/30 nm. (b) Sample A (period: 600 nm and Disc diameter 320 nm); (c) Sample B (period: 910 nm and disc diameter 360 nm).

**FIG. 2.** (Color online) Relative reflection spectra on two samples, reference sample: gold (100 nm thick)/silicon. (a) Reflection spectra measured using FTIR under normal incidence condition $(\theta=0^\circ)$, sample A (line with triangle symbol) and B (line with circle symbol); (b) Reflection ratio of $E_{TM}/E_{TE}$ on sample A (line with triangle symbol) and sample B (line with circle symbol) are measured using spectroscopic ellipsometer at incident angle of $(\theta=52^\circ)$.

**FIG.3.** Reflection ratio of $E_{TM}/E_{TE}$ on sample B under different incident angle $(\theta=48^\circ, 52^\circ, 56^\circ)$: (a-c) experimentally measured results (dashed line); (d-f) theoretically calculated results (solid line).

**FIG. 4.** (Color online) Both localized MPR mode and propagating SPP mode are excited by TM wave. $\vec{G}=2\pi/period$ is the reciprocal vector.

**FIG. 5.** (Color online) Experimental setup of the third harmonic generation. Near-Infrared was incident on the sample at angle $(\theta=52^\circ)$. The surface emission of THG (in-plane first order) was focused and then was imaged by CCD camera. The spectrum and intensity of third emission were recorded by Ocean Optics spectrometer and Newport power detector.

**FIG. 6.** (Color online) Bright field (a and b) and Dark field (c and d) imaging of third harmonic generation on sample A. Fundamental wavelength is: (a) 1740 nm; (b) 1800 nm; (c) 1650 nm; (d) 1790 nm respectively.

**FIG. 7.** (Color online) Spectra of third harmonic generation on sample A, fundamental wavelength: 1620 nm (line with square symbol); 1640 nm (line with triangle symbol).

**FIG. 8.** (Color online) Spectral resolved THG on sample A and B at incident angle of ($\theta=52^\circ$). (a) Experimental (line with circle symbol) and calculated (solid line) THG efficiency of sample A; (b) Experimental (line with circle symbol) and calculated THG efficiency of sample B (solid line).

**FIGURES**

**FIG. 1.**

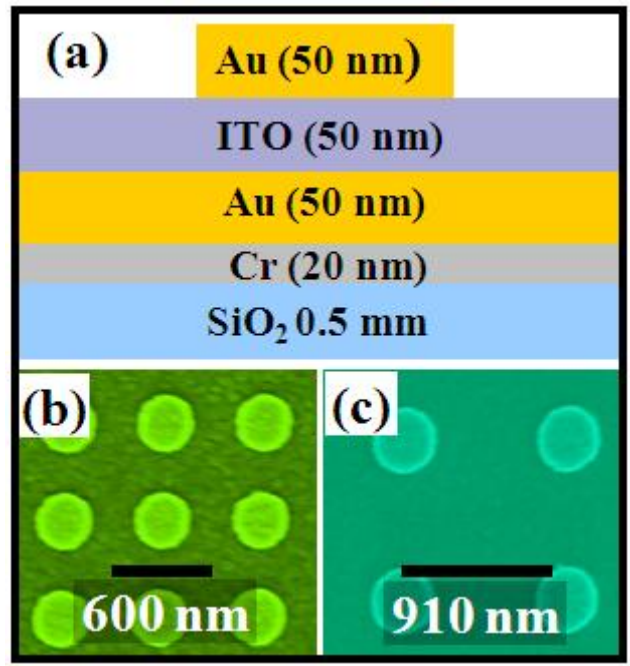

**FIG. 2.**

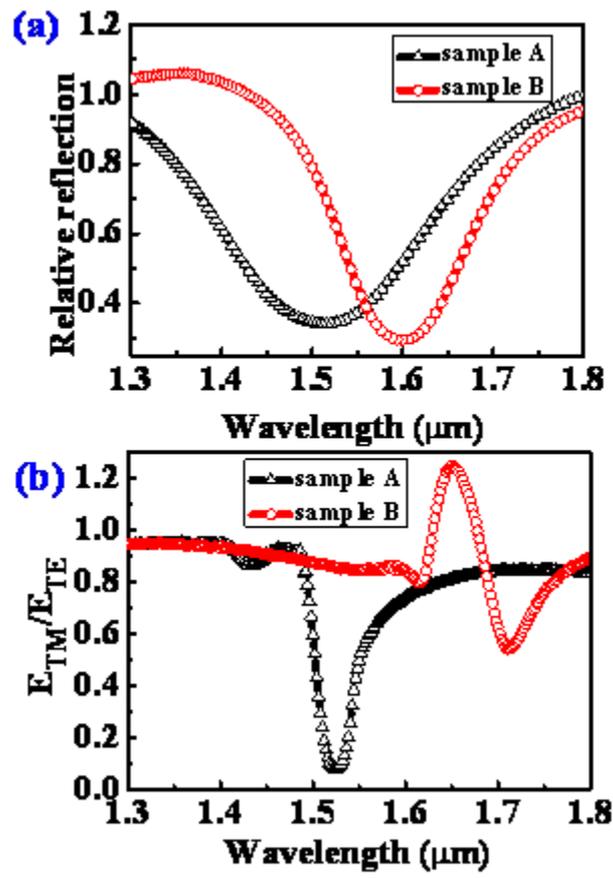

**FIG. 3.**

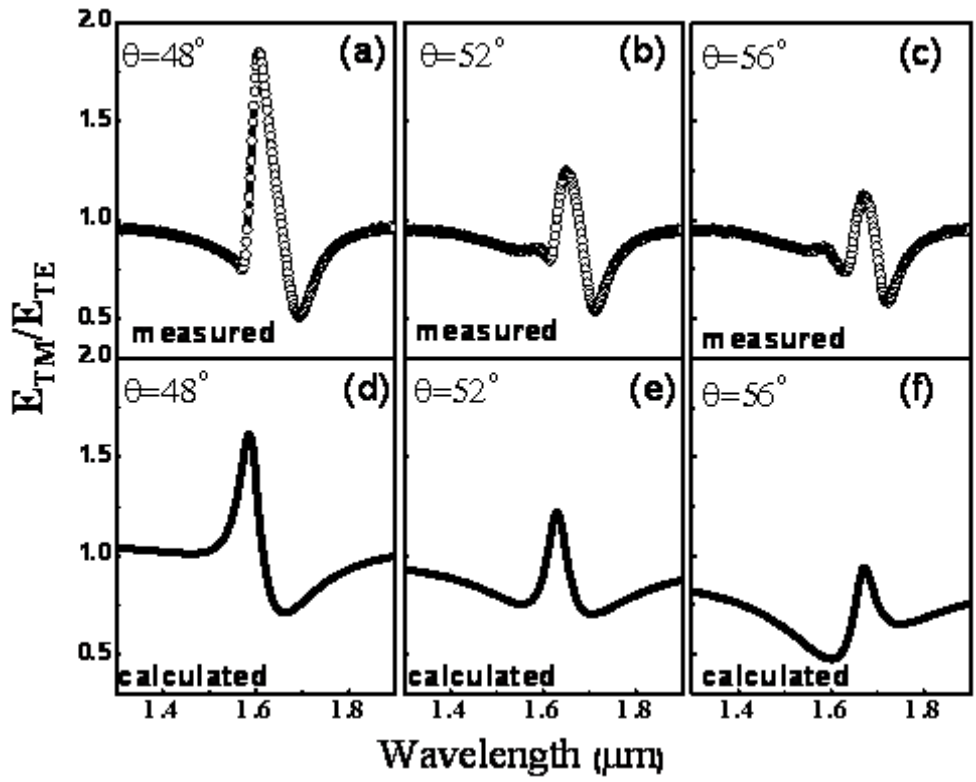

**FIG. 4.**

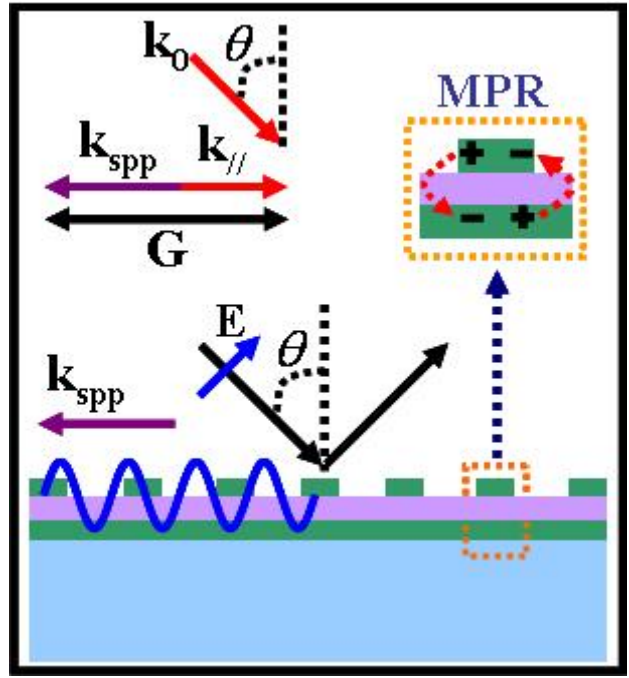

**FIG.5**

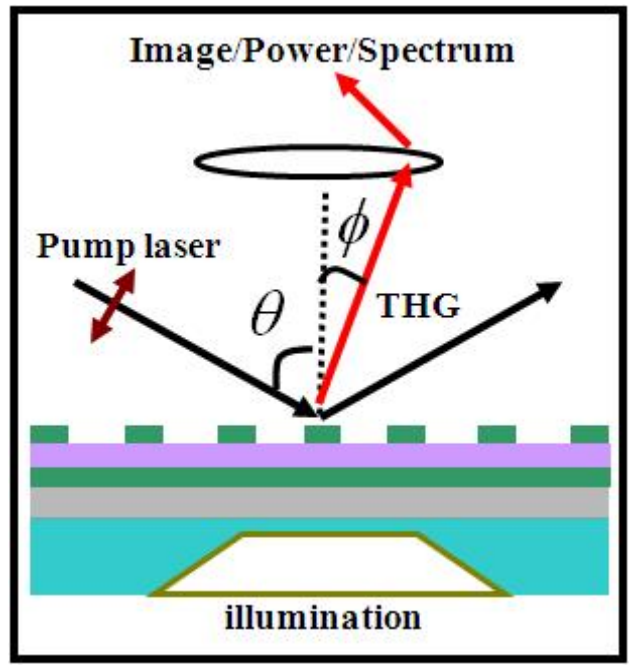

**FIG. 6**

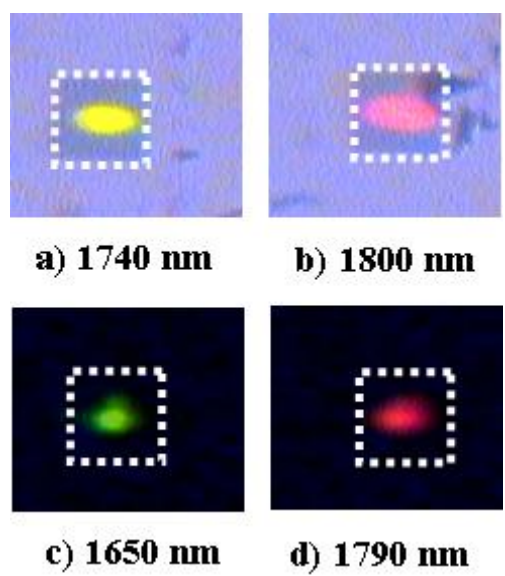

**FIG. 7**

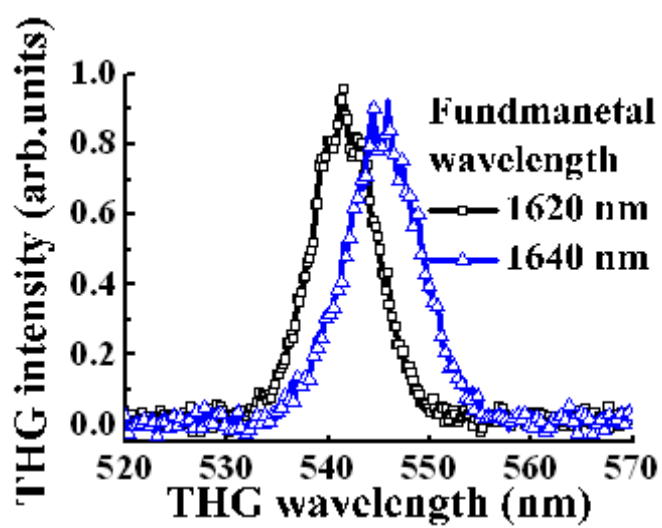

**FIG. 8**

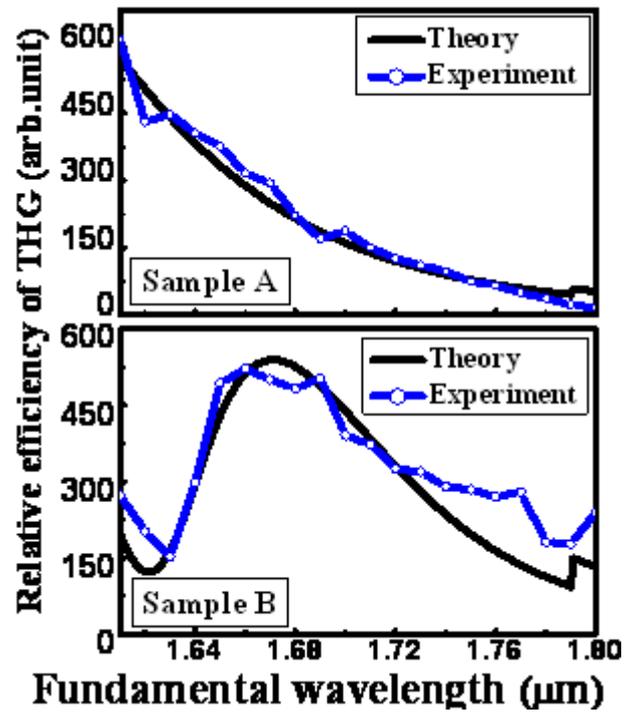